\documentclass[journal=psy]{for_review}
\makeatletter

\let\originalps@titlepage\ps@titlepage
\def\ps@titlepage{%
  \originalps@titlepage
  \let\@oddhead\@empty
  \let\@evenhead\@empty
}
\makeatother

\jname{}

\usepackage{graphicx}
\usepackage{caption}
\usepackage{placeins}
\usepackage{amsmath,amssymb,amsfonts}
\usepackage{float}
\usepackage{booktabs}
\usepackage{multirow}
\usepackage{threeparttable}
\usepackage{multirow}
\usepackage{graphicx}
\usepackage{tikz}
\usepackage{makecell}
\usetikzlibrary{arrows.meta, positioning, shapes.geometric, calc}
\usepackage{amsthm}
\usepackage{hyperref}
\theoremstyle{definition}
\newtheorem{example}{Example}
\defbibheading{bibliography}[\refname]{%
  \section*{#1}%
  \markboth{}{}%
}

\begin{document}

\begin{Frontmatter}

\title{Measuring Negotiation Traits with Collaborative Choice Data: A Two-Stage Latent Variable Model}
\date{July 2026}

\author[1,3]{Tongxin Zhang}
\author[1]{Yunxiao Chen}
\author[2]{Jiangang Hao}
\author[2]{Emily Kerzabi}
\author[2]{Patrick Kyllonen}
\author[1]{Irini Moustaki}
\author[3]{Tao Xin}

\address[1]{Department of Statistics, London School of Economics and Political Science}
\address[2]{Educational Testing Service, United States}
\address[3]{Collaborative Innovation Center of Assessment for Basic Education Quality, Beijing Normal University}

\maketitle
\begin{abstract}

Consider a collaborative decision-making setting in which team members with different reward structures need to make a joint decision. Repeated collaborative decisions can reflect individual differences in team members' negotiation ability. However, extracting such information is challenging because each observed response arises from a collaborative process involving multiple participants and their reward structures. In this paper, we propose a two-stage probabilistic measurement model for structured collaborative choice tasks. In the first stage, known member-specific rewards and latent participant negotiation traits jointly determine team-option utilities, from which a latent team choice is generated. In the second stage, individual responses are modeled conditional on the latent team choice, with possible deviations that depend on participants’ own rewards. Participant-level covariates can also be incorporated through a structural model. A simulation study shows satisfactory recovery of the participant traits and model parameters across different sample sizes and item sizes. An application to data from simulated collaborative negotiation tasks shows choice prediction accuracy well above the random-choice baseline and meaningful associations between the estimated traits and external criterion variables. The proposed framework provides a psychometric approach to measuring participant-level negotiation traits with collaborative choice data.
\end{abstract}

\noindent \textbf{Keywords:} collaborative problem solving; choice models; latent trait models; latent regression

\end{Frontmatter}

\section{Introduction}

Negotiation is an important 21st-century interpersonal skill in educational and workplace settings and a pervasive form of interpersonal decision making \citep{ThompsonEtAl2010,MartinRaughEtAl2020,JiangEtAl2023}. It typically has a mixed-motive structure, in which participants with potentially different interests must coordinate to reach an agreement while seeking outcomes that serve their own interests \citep{ThompsonEtAl2010,BoothbyEtAl2023}. Despite this interactive structure, individual outcomes and negotiation tactics have shown consistency across structurally similar negotiation tasks \citep{ElfenbeinEtAl2008,MartinRaughEtAl2020}. This consistency motivates the measurement of individual differences in negotiation.

Structured negotiation tasks used in educational assessment provide a controlled setting for such measurement. In these tasks, participants are asked to reach agreement on a common option while seeking to maximize their own total score. Across repeated items, they select from a common set of options, while the rewards associated with the same option may differ across participants. Recent collaborative problem-solving assessments have incorporated small-group negotiation tasks to investigate individual contributions to team performance \citep{KyllonenEtAl2023}. Participants must coordinate their selections to reach agreement on a common option \citep{MartinRaughEtAl2020,JiangEtAl2023}. These participant-specific rewards make individual interests explicit, and repeated choices across items may therefore provide information about individual differences in negotiation.

The repeated choices generated by these tasks naturally call for probabilistic choice modeling. In classical choice models, each alternative is associated with a latent value, termed utility, and its choice probability depends on its utility relative to those of the other alternatives \citep{Thurstone1927,Luce1959}. Extending this idea to psychometric measurement, Thurstonian IRT models express alternative utilities as functions of respondent traits, allowing these traits to be estimated from comparative responses \citep{MaydeuOlivaresBrown2010,BrownMaydeuOlivares2011}. These models use choice responses to measure individual differences, but remain centered on a single respondent. Economic models of collective decision making take a complementary approach by representing joint outcomes through the preferences or utilities of multiple members and a within-group decision mechanism \citep{ChiapporiMazzocco2017,ZhangKuwanoEtAl2009}. These models are generally developed to explain collective choice behavior rather than to estimate individual latent traits from repeated responses.

To combine individual-level measurement with a representation of collective choice, we propose a two-stage probabilistic measurement model for structured negotiation tasks. The two-stage formulation is related to item response tree models, which introduce latent intermediate decisions to represent the process underlying observed responses \citep{DeBoeckPartchev2012,JeonDeBoeck2016}. In the first stage, known member-specific rewards and latent negotiation traits jointly determine team--option utilities, with each participant's trait weighting that participant's reward profile. A latent team choice is then generated from these utilities through a Luce-type choice model. In the second stage, each member's submitted response is modeled conditional on the latent team choice. A member may follow the team choice or respond according to the member's own reward profile. 

Latent regression models introduce a structural component that relates systematic between-participant variation in a latent trait to observed person-level characteristics \citep{Mislevy1985}. Such models have been widely used in large-scale educational assessments to relate background characteristics to latent proficiency distributions \citep{AnderssonXin2021}. In the proposed framework, participant-level covariates may therefore be incorporated through a latent regression specification for the structural model of the negotiation trait.

The remainder of the paper is organized as follows. Section~2 describes the collaborative negotiation item data structure. Section~3 presents the proposed measurement model, its estimation, and its connections with related choice and measurement models. Section~4 evaluates parameter recovery through a simulation study. Section~5 applies the proposed model to data from computer-based negotiation tasks. Section~6 concludes with a discussion of the findings, limitations, and possible extensions.

\section{Data Structure of Collaborative Decision-making}

We consider collaborative negotiation items in which a team is required to reach an agreed choice from a finite set of options. A participant is a decision-making unit, such as an individual, an artificial intelligence agent, or a collective unit, whose internal interests can be represented by a single reward structure. For each item, the reward structure assigns a reward value to each available option for each member position in the team. The participants within a team aim to reach agreement on a set of items while making the agreed decisions as favorable as possible with respect to their own reward structures.

Let \(i = 1,\ldots,N\) index participants, and let \(j = 1,\ldots,J\) index collaborative negotiation items. Each item contains \(O\) available options, indexed by \(o = 1,\ldots,O\). Participants are organized into item-specific teams with a fixed team size \(M\). For item \(j\), let \(k = 1,\ldots,K_j\) index the teams formed for that item, and let \(m = 1,\ldots,M\) index the member positions within a team. A member position refers to the reward position assigned to a participant in a specific item-team context and is not assumed to be fixed for the same participant across items.

We denote by \(p_{jkm} \in \{1,\ldots,N\}\) the person-level index of the participant occupying member position \(m\) in team \(k\) on item \(j\). Thus, \((p_{jk1},\ldots,p_{jkM})\) gives the participants assigned to team \(k\) for item \(j\). The number of teams may vary across items, denoted by \(K_j\), as participants do not necessarily complete all items. This formulation also allows team composition to vary across items, so participants may work with different teammates and be assigned different member-position indices across items.

For each item \(j\), the known reward table specifies the rewards for all member positions and options. For modeling, the reward values are normalized by the maximum reward within the item to place the rewards on a common scale across items while preserving their relative magnitudes. Let $S_{jmo}\geq 0$ denote the resulting normalized reward value assigned to member position \(m\) if option \(o\) is selected. For member position \(m\), the row vector \((S_{jm1},\ldots,S_{jmO})\) gives the normalized reward vector across the \(O\) options. Larger values indicate more favorable outcomes for the corresponding member position.

The observed choice data are recorded at the item-team-member level. Let \(Y_{jkm} \in \{1,\ldots,O\}\) denote the option submitted by the participant occupying member position \(m\) in team \(k\) on item \(j\). If \(p_{jkm}=i\), then \(Y_{jkm}\) is the response submitted by participant \(i\) in that item-team context. In the model developed below, these submitted choices are treated as observed responses generated from an underlying team decision process.

\begin{example}[Reward Structure]
\label{ex:reward-table}

Consider an item \(j\) with \(M=4\) member positions and \(O=5\) options. The following raw reward table is taken from an item in the empirical application described in Section~\ref{sec:empirical-application}:
\[
\begin{array}{c|ccccc}
\hline
& \text{Option }1 & \text{Option }2 & \text{Option }3
& \text{Option }4 & \text{Option }5 \\
\hline
\text{Member position }1 & 0 & 75 & 100 & 25 & 50 \\
\text{Member position }2 & 75 & 0 & 25 & 100 & 50 \\
\text{Member position }3 & 25 & 100 & 75 & 0 & 50 \\
\text{Member position }4 & 100 & 25 & 0 & 75 & 50 \\
\hline
\end{array}
\]
\end{example}

In Example~\ref{ex:reward-table}, Suppose participant \(i\) occupies member position 2 in team \(k\) and submits Option 2. The participant-position mapping and observed response are therefore \(p_{jk2}=i\) and \(Y_{jk2}=2\), respectively. Under the normalization described above, the reward vector for member position 2 is
\[
(S_{j,2,1},\ldots,S_{j,2,5})
=
(0.75,0,0.25,1.00,0.50),
\]
and the reward corresponding to the submitted Option 2 is \(S_{j,2,2}=0\).

In addition to the choice and reward information, the data may also contain participant-level covariates. Let \(\mathbf{x}_i = (x_{i1},\ldots,x_{iD})^{\mathrm{T}}\) denote a vector of observed covariates for participant \(i\). These covariates may include background variables, prior characteristics, or behavioral indicators derived from the negotiation process. In the proposed model, \(\mathbf{x}_i\) is used to explain variation in the participant-level latent negotiation trait.

\section{Proposed Model}

We propose a team choice model for measuring participant-level negotiation traits from responses to collaborative negotiation items. The model comprises a two-stage measurement component and a structural component. The measurement component describes a two-stage process, with the formation of a latent team choice followed by the generation of individual responses conditional on that choice. The structural component extends the measurement model by incorporating participant-level covariates as additional information about the latent negotiation traits. This framework therefore accommodates both a baseline model based solely on the response data and a covariate-informed model in which the distribution of the latent traits depends on observed participant characteristics.

\subsection{Two-Stage Latent Negotiation Trait Measurement Model}

\begin{figure}[H]
    \centering
    \includegraphics[width=1\linewidth]{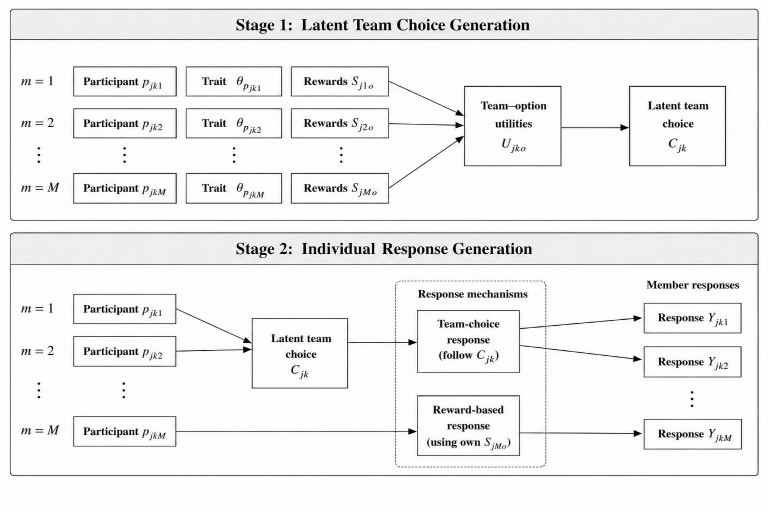}
    \caption{Overview of the two-stage response generation process.}
    \label{fig:two-stage-model}
\end{figure}

The proposed measurement model introduces a latent team-level choice to represent the collective decision underlying the observed member responses. Conditional on this latent choice, the responses submitted by individual team members are modeled separately and may differ from the team choice. Following the latent-process modeling approach used in item response tree models \citep{DeBoeckPartchev2012,JeonDeBoeck2016}, the measurement model is formulated as a two-stage process. Figure~\ref{fig:two-stage-model} provides an overview of the model structure.

\paragraph{Stage 1: Latent Team Choice Generation}

In the first stage, each participant is characterized by a strictly positive latent negotiation trait, denoted by $\theta_i\in(0,\infty)$. The latent trait determines the relative contribution of participant $i$'s reward profile to the utility of each possible team choice. For team $k$ on item $j$, the latent utility of option $o$ is specified as
\begin{equation}
U_{jko} = \beta_j \sum_{m=1}^{M} \theta_{p_{jkm}}S_{jmo}, 
\label{eq:team-option-utility}
\end{equation} 
where $\beta_j>0$ is a strictly positive item-specific choice sensitivity parameter. The term $\theta_{p_{jkm}}$ denotes the latent negotiation trait of the participant occupying member position $m$ in team $k$ on item $j$, where the index $p_{jkm}$ maps the member position to the corresponding $i$th participant. Within this model, $\theta_i$ represents the latent negotiation strength of participant $i$, operationalized as the relative extent to which the participant's reward profile influences the team's evaluation of the available options. In particular, a larger value of $\theta_i$ increases the contribution of participant $i$'s reward profile to the team utility. Thus, when team members favor different options, the reward profile of a participant with a larger latent negotiation trait has a stronger influence on the resulting team utilities. The trait therefore reflects a participant's relative influence in the team decision-making process, rather than merely the tendency to select options with high individual rewards. 

The utility $U_{jko}$ represents the team's latent evaluation of option $o$ following the negotiation process. It combines the known rewards of the team members, weighted by their respective latent negotiation traits. An option with a larger utility is regarded as more attractive to the team and is consequently more likely to become the latent team choice. The item parameter $\beta_j$ controls the degree to which differences in the weighted rewards distinguish the choice probabilities. When $\beta_j$ approaches zero, the probabilities of the $O$ options approach $1/O$. As $\beta_j$ increases, the choice distribution becomes increasingly concentrated on the option or options with the largest weighted rewards. 

Following Luce's choice model, the latent utilities induce a categorical choice distribution over the $O$ available options. For $t\in\{1,\ldots,O\}$, the probability of the latent team choice is given by 
\begin{equation} 
P\left( C_{jk}=t \mid \boldsymbol{\theta}_{jk}, \mathbf{S}_{j}, \beta_j \right) = \frac{\exp(U_{jkt})} {\sum_{o=1}^{O}\exp(U_{jko})}, 
\label{eq:team-choice-probability}
\end{equation} where $C_{jk}$ denotes the latent team choice for team $k$ on item $j$, $\boldsymbol{\theta}_{jk}
=(\theta_{p_{jk1}},\ldots,\theta_{p_{jkM}})$ denotes the vector of latent negotiation traits of the participants in team $k$ on item $j$, and $\mathbf{S}_{j}$ denotes the known reward table for item $j$ with elements $S_{jmo}$.

\paragraph{Stage 2: Individual Response Generation} 
In the second stage, the individual responses are generated conditionally on the latent team choice. Given $C_{jk}=t$, the response $Y_{jkm}$ submitted by the member in position $m$ is modeled as a mixture of directly following the latent team choice and responding according to the member's own rewards: 
\begin{equation} 
P(Y_{jkm}=y \mid C_{jk}=t) = (1-\pi_j)\mathbf{1}(y=t) + \pi_j P_{\mathrm{self}}(Y_{jkm}=y\mid \mathbf{S}_j,\alpha), 
\end{equation} 
for $y\in\{1,\ldots,O\}$, where $\mathbf{1}(y=t)$ equals one if the submitted response is the latent team choice and zero otherwise. Thus, the first component assigns probability $1-\pi_j$ to directly following the latent team choice. The parameter $\pi_j\in(0,1)$ denotes the item-specific probability that a response is generated through the self-oriented mechanism. Under the self-oriented mechanism, the response probability is determined by the rewards assigned to the corresponding member position: 
\begin{equation} 
P_{\mathrm{self}}(Y_{jkm}=y\mid \mathbf{S}_j,\alpha) = \frac{\exp\left(\alpha S_{jmy}\right)} {\sum_{o=1}^{O}\exp\left(\alpha S_{jmo}\right)}, \end{equation} 
where $\alpha\in(0,\infty)$ is a global reward sensitivity parameter in the self-oriented response distribution. As $\alpha$ approaches zero, the response probabilities approach $1/O$. Larger values of $\alpha$ produce a distribution that is increasingly concentrated on options providing higher rewards to the corresponding member position.

\addtocounter{example}{-1}
\begin{example}[Continued: Two-Stage Model Illustration]

For the item $j$ described in Example~\ref{ex:reward-table}, the normalized reward matrix used in the model is
\[ 
\mathbf{S}_{j} = 
\begin{pmatrix} 
0 & 0.75 & 1.00 & 0.25 & 0.50\\ 
0.75 & 0 & 0.25 & 1.00 & 0.50\\ 
0.25 & 1.00 & 0.75 & 0 & 0.50\\ 
1.00 & 0.25 & 0 & 0.75 & 0.50 
\end{pmatrix}. 
\]
Let the latent negotiation traits of the four participants occupying these member positions be
\[
\boldsymbol{\theta}_{jk}
=
(1.6,\,1.2,\,0.8,\,0.4),
\]
and let $\beta_j=2$. The aggregated trait-weighted contributions for the five options are
\[
\left(
\sum_{m=1}^{M}\theta_{p_{jkm}}S_{jm1},
\ldots,
\sum_{m=1}^{M}\theta_{p_{jkm}}S_{jm5}
\right)
=
(1.5,\,2.1,\,2.5,\,1.9,\,2.0).
\]
The corresponding latent team utilities are therefore
\[
\mathbf{U}_{jk}
=
(3.0,\,4.2,\,5.0,\,3.8,\,4.0),
\]
which yield the latent team choice probabilities
\[
\left(
P\left(C_{jk}=t
\mid
\boldsymbol{\theta}_{jk},\mathbf{S}_{j},\beta_j
\right)
\right)_{t=1}^{5}
=
(0.060,\,0.199,\,0.444,\,0.134,\,0.163).
\]
Option 3 has the highest probability of becoming the latent team choice. In this example, the participant in the first position has the largest latent negotiation trait ($\theta_{p_{jk1}}=1.6$) and receives the highest reward under Option 3 (${S}_{j,1,3} = 1.00$), thereby contributing substantially to the utility of Option 3.

For the second stage, suppose that the realized latent team choice is $C_{jk}=3$. Consider the participant in the second position, whose rewards across the five options are
\[
(S_{j,2,1},S_{j,2,2},S_{j,2,3},S_{j,2,4},S_{j,2,5})
=
(0.75,\,0,\,0.25,\,1.00,\,0.50).
\]
This participant receives the highest individual reward under Option 4 rather than the latent team choice. Let $\alpha=2$ and $\pi_j=0.25$. The self-oriented response probabilities for this participant are
\[ 
\left( P_{\mathrm{self}}(Y_{jk2}=1\mid \mathbf{S}_j,\alpha ), \ldots, P_{\mathrm{self}}(Y_{jk2}=5\mid \mathbf{S}_j,\alpha) \right) 
=
(0.260,\,0.058,\,0.096,\,0.429,\,0.158). 
\]
Combining these probabilities with the probability of directly following the latent team choice gives 
\[ 
\left( P(Y_{jk2}=1\mid C_{jk}=3), \ldots, P(Y_{jk2}=5\mid C_{jk}=3) \right) 
=(
0.065,\,0.015,\,0.774,\,0.107,\,0.039). 
\] 
Although the participant's individual rewards favor Option 4, Option 3 remains the most probable submitted response because the participant directly follows the latent team choice with probability $1-\pi_j=0.75$. The remaining probability is distributed across the options according to the participant's own rewards.

\end{example}

\subsection{Structural Model for the Latent Negotiation Trait}

We specify a structural model to characterize the distribution of the participant-level latent negotiation trait. Two specifications are considered. The baseline specification models the latent trait distribution without participant-level covariates, whereas the latent regression specification incorporates observed participant-level covariates.

\paragraph{Baseline Specification}

In the baseline specification, the distribution of the latent negotiation trait is modeled without participant-level covariates. Specifically,
\begin{equation}
\log(\theta_i)
\sim
N\left(0,\sigma_\theta^2\right),
\end{equation}
where the mean of the log latent trait is fixed at zero. The variance parameter $\sigma_\theta^2>0$ governs the between-participant heterogeneity in the log latent negotiation trait and is estimated jointly with the other model parameters.

\paragraph{Latent Regression Specification}

Participant-level covariates may contain information about individual differences in the negotiation trait beyond that provided by the observed choice responses. We therefore use a latent regression specification to incorporate this additional information into the structural model \citep{Mislevy1985,AnderssonXin2021}. Specifically, the conditional mean of the log latent negotiation trait is related to the observed covariates. Let $\mathbf{x}_i$ denote the centered covariate vector for participant $i$. The latent regression model is specified as
\begin{equation}
\log(\theta_i)\mid\mathbf{x}_i
\sim
N\left(
\boldsymbol{\gamma}^{\mathrm{T}}\mathbf{x}_i,
\sigma_\theta^2
\right),
\end{equation}
where $\boldsymbol{\gamma}$ denotes the corresponding vector of regression coefficients. The covariates enter only through the conditional mean of $\log(\theta_i)$ and provide additional information about individual differences in the latent negotiation trait. The variance parameter $\sigma_\theta^2$ represents the residual between-participant variation in the log latent negotiation trait after accounting for the observed covariates and is estimated jointly with the other model parameters.

\begin{figure}[htbp]
    \centering
    \includegraphics[width=0.7\textwidth]{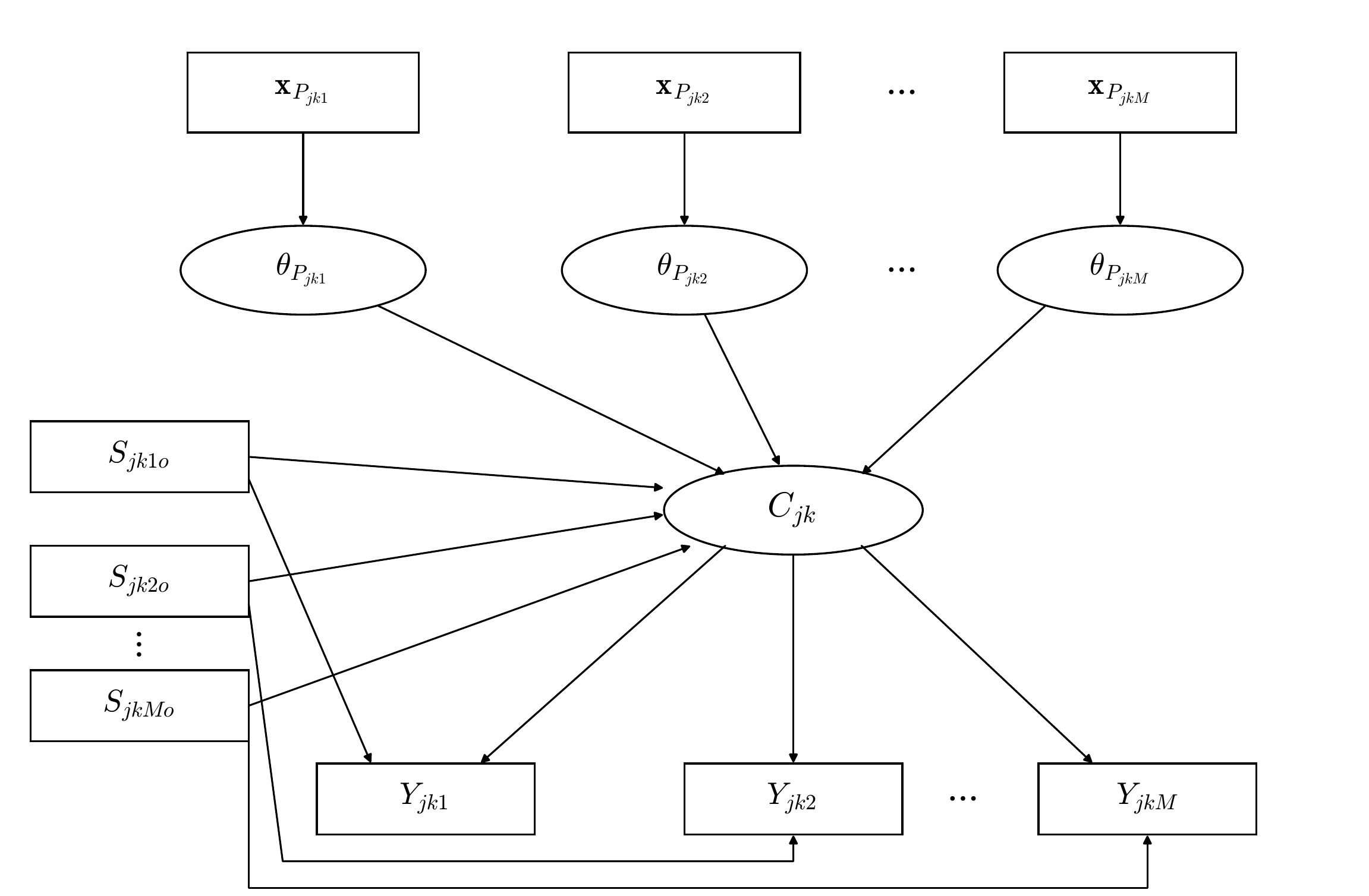}
    \caption{Graphical representation of the proposed team choice model with latent regression}
    \label{fig:model_structure}
\begin{flushleft}
\footnotesize
\textit{Note.} Rectangles represent observed variables, ellipses represent latent variables, and arrows indicate the conditional relationships specified in the model.
\end{flushleft}
\end{figure}

The relationships among the observed variables, latent variables, and reward structure specified in the measurement and structural components are summarized in Figure~\ref{fig:model_structure}.

\subsection{Inference}

The likelihood is constructed at the item--team level, consistent with the two-stage generation of the observed member responses. For team $k$ on item $j$, let
\[
\mathbf{Y}_{jk}
=
(Y_{jk1},\ldots,Y_{jkM})
\]
denote the response vector, and let
\[
\mathbf{y}_{jk}
=
(y_{jk1},\ldots,y_{jkM})
\]
denote its observed realization. Let
\[
\boldsymbol{\theta}_{jk}
=
\left(
\theta_{p_{jk1}},\ldots,\theta_{p_{jkM}}
\right)
\]
denote the latent negotiation traits of the participants in that team. Conditional on $\boldsymbol{\theta}_{jk}$, the likelihood contribution for team $k$ on item $j$ is
\begin{equation}
\begin{aligned}
L_{jk}
\left(
\boldsymbol{\theta}_{jk},
\beta_j,
\pi_j,
\alpha
\right)
&=
P\left(
\mathbf{Y}_{jk}
=
\mathbf{y}_{jk}
\mid
\boldsymbol{\theta}_{jk},
\beta_j,
\pi_j,
\alpha
\right)\\
&=
\sum_{t=1}^{O}
P\left(
C_{jk}=t
\mid
\boldsymbol{\theta}_{jk},
\beta_j
\right)
\prod_{m=1}^{M}
P\left(
Y_{jkm}
\mid
C_{jk}=t,
\pi_j,
\alpha
\right).
\end{aligned}
\end{equation}
The summation marginalizes over the unobserved team choice $C_{jk}$, whereas the product over member positions follows from the conditional independence of the individual responses given the latent team choice. Conditioning on the known reward structure is omitted from the notation for simplicity.

Let $\mathbf{y}$ denote the collection of all observed member responses. Conditional on the participant-level latent negotiation traits, the response vectors from different item--team pairs are assumed to be independent. The conditional likelihood of all observed responses is therefore
\begin{equation}
L\left(
\mathbf{y}
\mid
\boldsymbol{\theta},
\boldsymbol{\beta},
\boldsymbol{\pi},
\alpha
\right)
=
\prod_{j=1}^{J}
\prod_{k=1}^{K_j}
L_{jk}
\left(
\boldsymbol{\theta}_{jk},
\beta_j,
\pi_j,
\alpha
\right),
\end{equation}
where
\[
\qquad
\boldsymbol{\theta}
=
(\theta_1,\ldots,\theta_N),
\boldsymbol{\beta}
=
(\beta_1,\ldots,\beta_J),\boldsymbol{\pi}
=
(\pi_1,\ldots,\pi_J).
\]
Although team composition may vary across items, responses from the same participant are linked through the common latent negotiation trait $\theta_i$.

The marginal likelihood is obtained by integrating the conditional likelihood over the structural distribution of the participant-level latent negotiation traits. Under the latent regression specification,
\begin{equation}
\begin{aligned}
L_{\mathrm{marg}}
\left(
\mathbf{y}
\mid
\mathbf{X},
\boldsymbol{\beta},
\boldsymbol{\pi},
\alpha,
\boldsymbol{\gamma},
\sigma_\theta^2
\right)
=
\int_{(0,\infty)^N}
&
L\left(
\mathbf{y}
\mid
\boldsymbol{\theta},
\boldsymbol{\beta},
\boldsymbol{\pi},
\alpha
\right)\\
&\times
\prod_{i=1}^{N}
f_\theta
\left(
\theta_i
\mid
\mathbf{x}_i,
\boldsymbol{\gamma},
\sigma_\theta^2
\right)
\,d\boldsymbol{\theta}.
\end{aligned}
\end{equation}
The integral is \(N\)-dimensional because it is taken jointly over the \(N\) participant-level latent negotiation traits. Although the structural density factorizes across participants, the conditional likelihood generally does not. Each item--team likelihood contribution depends jointly on the latent negotiation traits of all team members, and the same participant may appear in multiple teams across items. Therefore, the marginal likelihood cannot generally be decomposed into a product of one-dimensional integrals.

Here, $f_\theta(\cdot)$ denotes the log-normal density implied by the structural model. For the baseline specification, the covariates and regression coefficients are omitted, and
$f_\theta(\theta_i,\sigma_\theta^2)$ denotes the log-normal density with $\log(\theta_i)\sim N(0,\sigma_\theta^2)$. In both specifications, $\sigma_\theta^2$ represents the residual variance of the log-transformed latent negotiation traits and is estimated jointly with the remaining model parameters.

The model is estimated within a Bayesian framework. Prior distributions are assigned to $\boldsymbol{\beta}$, $\boldsymbol{\pi}$, $\alpha$, $\boldsymbol{\gamma}$, and $\sigma_\theta$, while the distribution of the participant-level latent negotiation traits is specified by the corresponding structural model. The prior specifications used in the simulation and empirical studies are provided in Appendix~\ref{app:estimation}. Posterior inference is conducted using Markov chain Monte Carlo methods, including conjugate Gibbs updates and adaptive random-walk Metropolis--Hastings updates, with posterior computation implemented in R using the \texttt{nimble} package.

\subsection{Connections with Related Models}

The proposed model is related to three lines of work: classical choice
models, item response tree models, and Thurstonian item response theory (IRT) models.

\paragraph{Connection with Classical Choice Models}

Classical choice models relate observed choices to latent utilities of the alternatives. Thurstone's model represents paired judgments
through differences in latent utility \citep{Thurstone1927}. For choices among multiple alternatives, Luce's model assigns each alternative a positive scale value and defines its choice probability relative to the scale values of all available alternatives
\citep{Luce1959}. Let \(v_o>0\) denote the scale value of alternative \(o\). The probability of choosing alternative \(t\) is
\begin{equation}
P(C=t)
=
\frac{v_t}
{\sum_{o=1}^{O}v_o}
=
\frac{\exp(U_t)}
{\sum_{o=1}^{O}\exp(U_o)},
\end{equation}
where \(U_o=\log v_o\) denotes the utility of alternative \(o\). The Plackett--Luce model applies the same choice rule sequentially to ranking data \citep{Plackett1975}.

The first stage of the proposed model adopts the same Luce-type choice probability; see Equations~\eqref{eq:team-option-utility} and \eqref{eq:team-choice-probability}. The utility specification, however, differs from that of classical choice models. Utility is defined at the team--option level and combines the known reward profiles and latent negotiation traits of multiple participants. Moreover, \(C_{jk}\) is a latent team choice rather than an observed response.

\paragraph{Connection with Multistage Latent Response Models}

The two-stage structure of the proposed model is related to multistage latent response models, in which observed responses are generated through unobserved intermediate processes or response states. Item response tree (IRTree) models represent an observed categorical response as the outcome of a sequence of latent decisions \citep{DeBoeckPartchev2012,JeonDeBoeck2016}. Related latent response models introduce an unobserved response state that determines the mechanism generating the observed data. For example, \citet{EstherEtAl2020} introduced an item-level engagement state to distinguish engaged responding from disengagement manifested in guessing or item nonresponse, whereas \citet{EstherEtAl2022} introduced a screen-level attentiveness state to distinguish attentive responding from careless or insufficient-effort responding using both responses and response times.

The proposed model shares this general feature by introducing a latent team choice \(C_{jk}\) between the team--option utilities and the observed member responses \(Y_{jkm}\). The latent team choice represents an intermediate stage in the generation of individual responses. Its interpretation, however, differs from those of the intermediate variables in existing models. IRTree and latent response models typically introduce respondent-level decisions or response states to explain the response of a single respondent. In the proposed model, the intermediate variable is instead a team-level choice shared by all members of the team, while the observed responses remain member-specific.

\paragraph{Connection with Thurstonian IRT Models}

Thurstonian IRT models extend Thurstone's comparative judgment framework to paired-comparison, ranking, and forced-choice data by expressing alternative-specific latent utilities as functions of respondent traits and item parameters \citep{MaydeuOlivaresBockenholt2005,
MaydeuOlivaresBrown2010,
BrownMaydeuOlivares2011}.

Both frameworks link latent traits to choice through option utilities. In Thurstonian IRT models, the utilities describe alternatives
evaluated by a single respondent. In the proposed model, utility is defined for a team--option pair. The rewards \(S_{jkmo}\) are known from the item design, and each participant's trait weights the corresponding reward profile in the team--option utility.

\section{Simulation Study}

We conducted a simulation study to evaluate the recovery of the model parameters and participant-level latent negotiation traits under different sample sizes and number of items. The proposed framework includes both baseline and covariate-informed specifications. These two models were fitted to the same simulated datasets to examine whether incorporating participant-level covariates improves the estimation of latent traits.

\paragraph{Simulation Setting}

We varied the sample size and the number of items, each at four levels: \(N\in\{120,300,600,1200\}\) and \(J\in\{10,20,30,50\}\), yielding 16 simulation conditions. The team size and the number of response options were fixed at \(M=4\) and \(O=5\), respectively. The \(J\) items were divided equally between two task blocks, with \(J/2\) items in each block. Participants were randomly assigned to teams separately within each block, and the same team completed all items within a block. Overall, each simulated dataset contained \(NJ/4\) item--team observations.

The reward structures followed the design of the collaborative negotiation tasks described in previous assessment studies \citep{KyllonenEtAl2023}. Each reward table was normalized to have a maximum value of one, and the 10 reward tables were repeated as needed to generate the specified number of items.

Fourteen participant-level covariates were generated and standardized to have mean zero and variance one. Their regression coefficients were specified as \(\boldsymbol{\gamma}=(0,.30,0,-.10,0,.20,0,
0,.30,0,-.10,0,.20,0)^{\mathrm{T}}\) including positive, negative, and null effects. The participant level latent traits were generated according to
\[
\log\theta_i
=
\boldsymbol{\gamma}^{\mathrm{T}}\mathbf{x}_i
+
\varepsilon_i,
\qquad
\varepsilon_i\sim N(0,0.70^2).
\]

The item-specific parameters were generated with \(\log\beta_j\sim N(0.50,0.50^2)\) and \(\pi_j\sim\operatorname{Beta}(2,8)\), and the global reward sensitivity
parameter was fixed at \(\alpha=5\). Conditional on the generated parameters and the reward structures, latent team choices and individual responses were generated according to the proposed two-stage measurement model.

Each simulated dataset was analyzed using the two model specifications described in the proposed model section. Data were generated under the latent regression specification, whereas the baseline model omitted participant-level covariates while retaining the same measurement component. Posterior means were used as point estimates. For each simulation condition, 50 independent replications were generated and analyzed under both specifications.

Recovery was evaluated using root mean squared error (RMSE) for all model parameters and the correlation between the true and estimated latent negotiation traits. For each replication, RMSEs were calculated across participants for the latent negotiation traits, across items for the item-specific parameters, and across covariates for the regression coefficients. For latent negotiation traits, RMSE was calculated on both the original and logarithmic scales. The reported results summarize the mean recovery measures across replications with 95\% Monte Carlo confidence intervals.

\paragraph{Results}

\begin{figure}[H]
    \centering
    \includegraphics[
        width=0.7\textwidth,
    ]{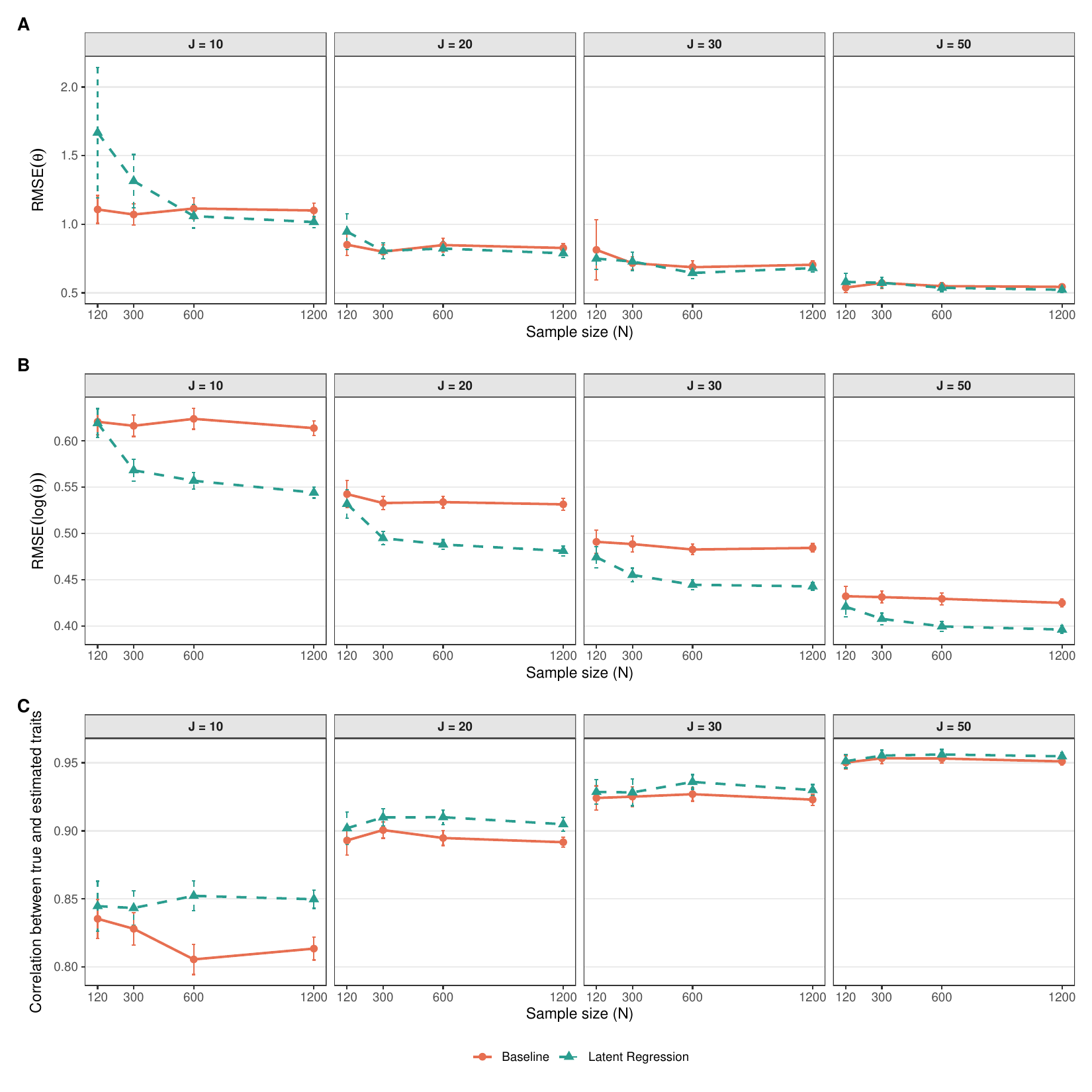}
    
    \caption{Recovery of the latent negotiation traits across simulation conditions.}
    \label{fig:theta_recovery}
    
    \begin{minipage}{\textwidth}
        \footnotesize
        \textit{Note.}
        Panels A--C show the root mean squared error of the latent negotiation traits on the original scale, the root mean squared error on the logarithmic scale, and the correlation between the true and estimated latent negotiation traits, respectively. Points represent means across 50 replications, and error bars indicate 95\% Monte Carlo confidence intervals. 
    \end{minipage}
\end{figure}

The recovery of the latent negotiation traits is summarized in Figure~\ref{fig:theta_recovery}. Recovery improved with larger sample sizes and longer tests, as indicated by lower RMSEs on both the original and logarithmic scales and higher correlations between the true and estimated traits. Under the largest simulation condition (\(N=1200, J=50\)), the RMSEs on the original and logarithmic scales were around 0.5 and 0.4, respectively, and the correlations between the true and estimated traits exceeded 0.95 for both model specifications. The correlations remained above 0.80 across all conditions. The effects of sample size and the number of items were consistent across the two model specifications.

The latent regression model achieved lower RMSEs on the logarithmic scale (Panel B) and higher correlations between the true and estimated traits (Panel C) across most conditions, suggesting that incorporating participant-level covariates improved trait recovery. The improvement was less evident on the original \(\theta\) scale (Panel A), where the two specifications showed comparable RMSEs across most conditions. Under \(J=10\) and the two smallest sample sizes (\(N=120\) and \(N=300\)), the latent regression model yielded larger RMSEs on the original \(\theta\) scale than the baseline model. These conditions also showed greater variability in RMSEs across replications, which may reflect the additional uncertainty introduced by estimating the regression coefficients with limited participant information and recovering latent negotiation traits with limited response information from short tests. The greater variability observed on the original \(\theta\) scale may also reflect the right-skewness of its lognormal distribution, where a small number of large trait values contributed disproportionately to the squared error.

\begin{figure}[H]
    \centering
    \includegraphics[
        width=0.65\textwidth,
    ]{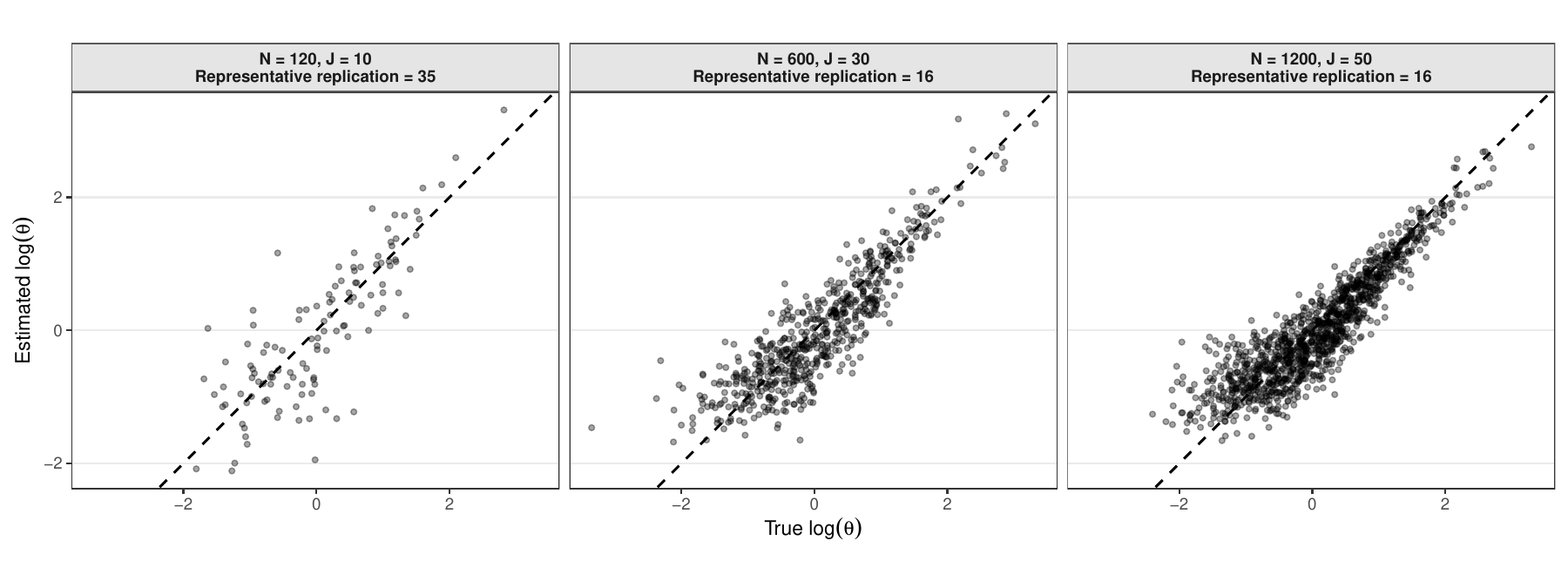}
    
    \caption{True and estimated log-transformed latent negotiation traits under representative simulation conditions.}
    \label{fig:theta_scatter}
    
    \begin{minipage}{\textwidth}
        \footnotesize
        \textit{Note.}
        For each condition, the displayed replication has a log-scale trait RMSE closest to the median RMSE across replications. The dashed line
        denotes the identity line.
    \end{minipage}
\end{figure}

The representative scatter plots in Figure~\ref{fig:theta_scatter} further illustrate the agreement between the true and estimated latent negotiation traits on the logarithmic scale, with estimates showing closer alignment with the identity line under larger sample sizes and longer tests.

\begin{figure}[!htbp]
    \centering
    \includegraphics[
        width=0.7\textwidth,
    ]{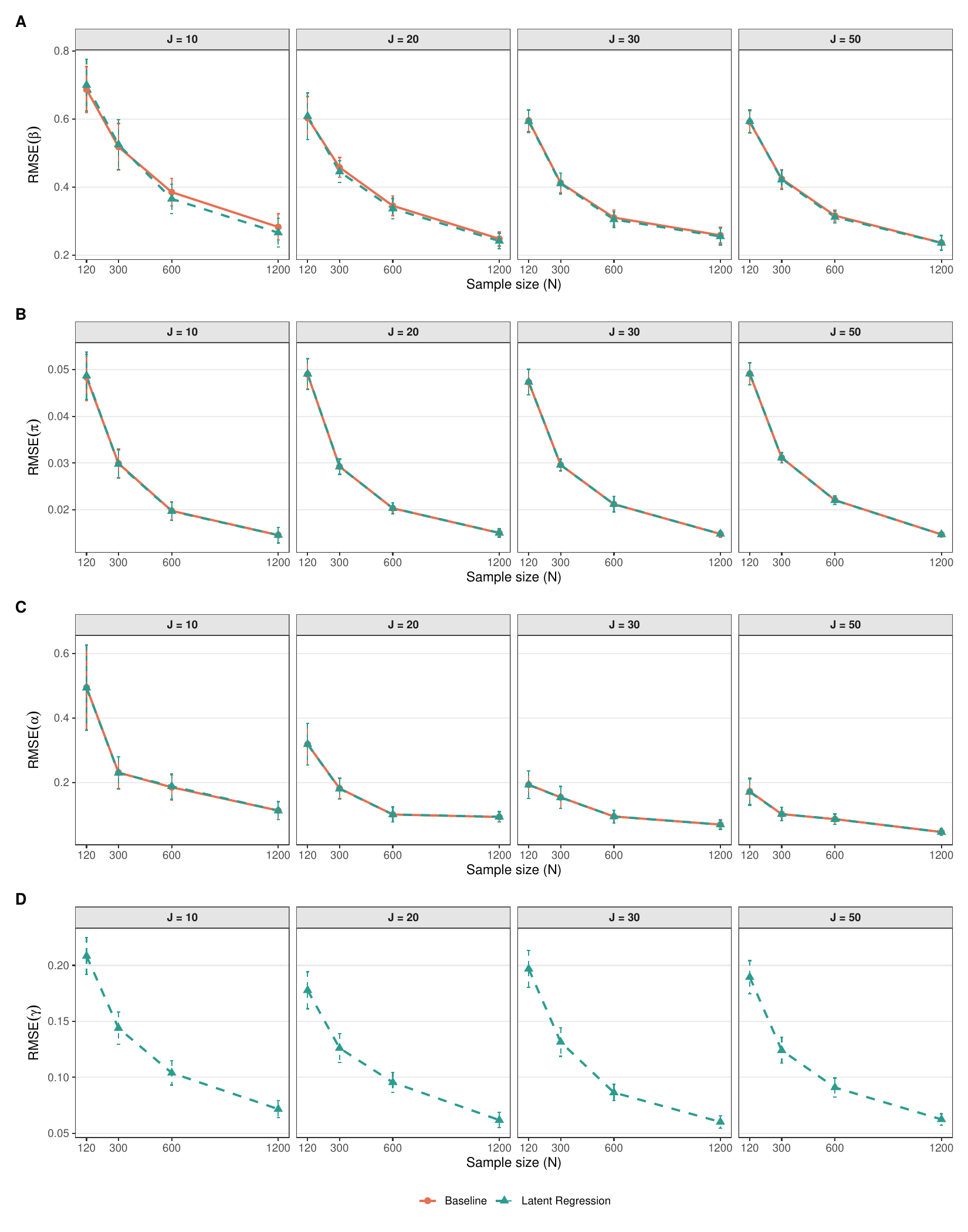}
    
    \caption{Recovery of the measurement and structural parameters across simulation conditions.}
    \label{fig:parameter_recovery}
    
    \begin{minipage}{\textwidth}
        \footnotesize
        \textit{Note.}
        Panels A--D show the root mean squared errors of the item-specific reward sensitivity parameters \(\beta_j\), the item-specific self-oriented response probability parameters \(\pi_j\), the global reward sensitivity parameter \(\alpha\), and the regression coefficients \(\boldsymbol{\gamma}\), respectively. Points represent means across replications, and error bars indicate 95\% Monte Carlo confidence intervals. The regression coefficients $\boldsymbol{\gamma}$ are included only in the latent regression model.
    \end{minipage}
\end{figure}

The recovery of the remaining model parameters is shown in Figure~\ref{fig:parameter_recovery}. The RMSEs of the measurement parameters generally decreased with larger sample sizes. The effect of the number of items varied across parameters. For the item-specific reward sensitivity parameters \(\beta_j\) and the item-specific self-oriented response probability parameters \(\pi_j\), the number of items had limited effects on RMSE values, although increasing the number of items tended to reduce the Monte Carlo variability across replications. For the global reward sensitivity parameter \(\alpha\), recovery improved as both the sample size and the number of items increased. The two model specifications showed nearly identical recovery patterns for the measurement parameters, indicating that incorporating participant-level covariates did not substantially affect the estimation of the measurement component.

The regression coefficients \(\boldsymbol{\gamma}\), included only in the latent regression model, were recovered more accurately as the sample size increased. Their recovery showed little dependence on the number of items, suggesting that the estimation of participant-level regression effects was primarily driven by the number of participants. Together with the recovery results for the latent negotiation traits, these findings show that incorporating participant-level covariates improved the recovery of latent negotiation traits while preserving the estimation of the measurement
component.

\FloatBarrier

\section{Case Study}
\label{sec:empirical-application}
\subsection{Data and Analytic Procedure}
\paragraph{Background}
The empirical analysis used data from two parallel collaborative negotiation tasks developed and administered by Educational Testing Service (ETS) \citep{KyllonenEtAl2023, HaoEtAl2025}. The data were collected through Prolific between 2021 and 2023 and also included the Big Five Inventory--2 Short Form (BFI-2-S), a cognitive assessment, and the STEM-B assessment. Participants completed each task in randomly formed four-person teams, with teammates reassigned across tasks. We refer to the Fundraiser Negotiation and Company-wide Presentation tasks as Task Na and Task Nb, respectively.

\begin{figure}[htbp]
    \centering
    \includegraphics[width=0.7\textwidth]
    {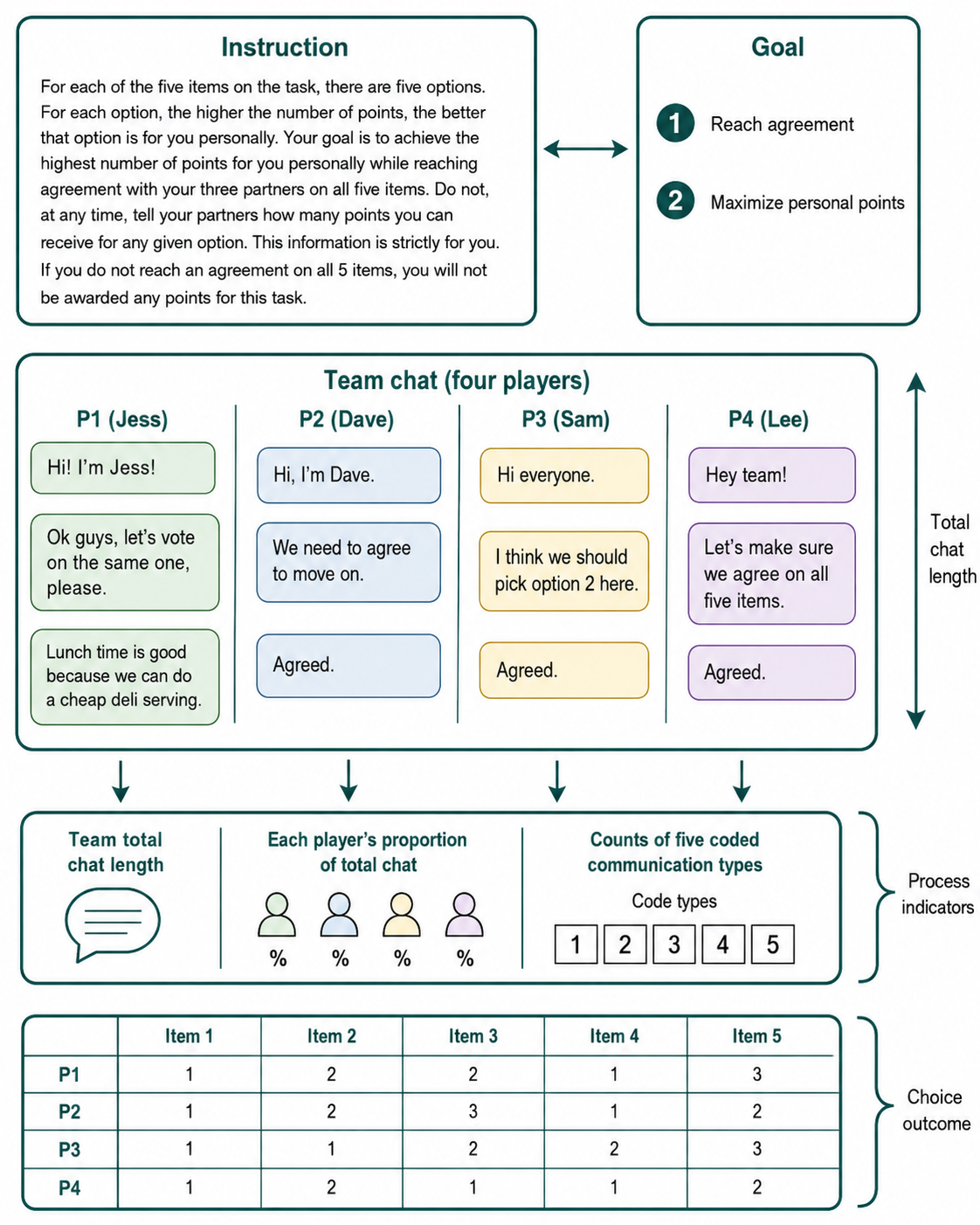}
    \caption{Task Procedure}
    \label{fig:task-procedure}
\end{figure}

The task procedure is shown in Figure~\ref{fig:task-procedure}. Participants first reviewed the instructions and their role-specific reward table, then negotiated through a chat interface while revising their item choices, and finally submitted responses to the five items.

Teams were instructed to reach agreement on all items while each participant sought to maximize their own points. The tasks included asymmetric, compatible, and symmetric items (see Table~\ref{tab:reward_tables}). Their parallel design balanced the score scale across player roles: over the five items, each role had the same maximum total score of 2,200, minimum total score of 0, and average total score of 1,100.

\begin{table}[htbp]
\centering
\caption{Reward tables for the two collaborative negotiation tasks}
\label{tab:reward_tables}
\scriptsize

\textbf{Panel A. Task Na: Fundraiser Negotiation}

\resizebox{\textwidth}{!}{%
\begin{tabular}{c cccc cccc cccc cccc cccc}
\toprule
\multirow{2}{*}{Option} 
& \multicolumn{4}{c}{Item 1: Asymmetric} 
& \multicolumn{4}{c}{Item 2: Compatible} 
& \multicolumn{4}{c}{Item 3: Asymmetric} 
& \multicolumn{4}{c}{Item 4: Symmetric} 
& \multicolumn{4}{c}{Item 5: Symmetric} \\
\cmidrule(lr){2-5} \cmidrule(lr){6-9} \cmidrule(lr){10-13} \cmidrule(lr){14-17} \cmidrule(lr){18-21}
& P1 & P2 & P3 & P4
& P1 & P2 & P3 & P4
& P1 & P2 & P3 & P4
& P1 & P2 & P3 & P4
& P1 & P2 & P3 & P4 \\
\midrule
1 & 150 & 50 & 0   & 12.5 & 50 & 50 & 50 & 50 & 50   & 150 & 12.5 & 0   & 50 & 50 & 50  & 50  & 0   & 75 & 25  & 100 \\
2 & 200 & 37.5 & 50 & 0    & 100 & 100 & 100 & 100 & 12.5 & 0   & 50   & 150 & 25 & 0  & 100 & 75  & 75  & 0  & 100 & 25 \\
3 & 50  & 0   & 200 & 37.5 & 75 & 25 & 25 & 75 & 37.5 & 200 & 0    & 50  & 0  & 25 & 75  & 100 & 100 & 25 & 75  & 0 \\
4 & 100 & 25  & 100 & 25   & 0  & 0  & 0  & 0  & 25   & 100 & 25   & 100 & 75 & 100 & 0   & 25  & 25  & 100 & 0   & 75 \\
5 & 0   & 12.5 & 150 & 50  & 25 & 75 & 75 & 25 & 0    & 50  & 37.5 & 200 & 100 & 75 & 25 & 0   & 50  & 50 & 50  & 50 \\
\bottomrule
\end{tabular}%
}

\vspace{1em}

\textbf{Panel B. Task Nb: Company-wide Presentation}

\resizebox{\textwidth}{!}{%
\begin{tabular}{c cccc cccc cccc cccc cccc}
\toprule
\multirow{2}{*}{Option} 
& \multicolumn{4}{c}{Item 1: Asymmetric} 
& \multicolumn{4}{c}{Item 2: Symmetric} 
& \multicolumn{4}{c}{Item 3: Compatible} 
& \multicolumn{4}{c}{Item 4: Asymmetric} 
& \multicolumn{4}{c}{Item 5: Symmetric} \\
\cmidrule(lr){2-5} \cmidrule(lr){6-9} \cmidrule(lr){10-13} \cmidrule(lr){14-17} \cmidrule(lr){18-21}
& P1 & P2 & P3 & P4
& P1 & P2 & P3 & P4
& P1 & P2 & P3 & P4
& P1 & P2 & P3 & P4
& P1 & P2 & P3 & P4 \\
\midrule
1 & 200 & 37.5 & 50  & 0   & 50  & 50  & 50  & 50  & 0   & 0   & 0   & 0   & 50  & 150 & 12.5 & 0   & 0   & 75  & 25  & 100 \\
2 & 100 & 25   & 100 & 25  & 25  & 0   & 100 & 75  & 75  & 25  & 25  & 75  & 37.5 & 200 & 0    & 50  & 25  & 100 & 0   & 75 \\
3 & 150 & 50   & 0   & 12.5& 75  & 100 & 0   & 25  & 50  & 50  & 50  & 50  & 12.5 & 0   & 50   & 150 & 50  & 50  & 50  & 50 \\
4 & 0   & 12.5 & 150 & 50  & 100 & 75  & 25  & 0   & 100 & 100 & 100 & 100 & 0    & 50  & 37.5 & 200 & 75  & 0   & 100 & 25 \\
5 & 50  & 0    & 200 & 37.5& 0   & 25  & 75  & 100 & 25  & 75  & 75  & 25  & 25   & 100 & 25   & 100 & 100 & 25  & 75  & 0 \\
\bottomrule
\end{tabular}%
}

\begin{flushleft}
\footnotesize
\textit{Note.} Task Na required teams to negotiate five decisions concerning the type of event, the beneficiary, the fundraising goal, the time for a planning meeting, and the timing of the event within the semester. Task Nb required teams to negotiate five decisions concerning the presentation location, delivery method, source of feedback, preparation meeting time, and presentation length.
\end{flushleft}

\end{table}

\paragraph{A summary of data}

The dataset contains detailed process and outcome data from the collaborative negotiation tasks. The process data consist of the chat messages generated in the chat box during team negotiation. Following \citet{HaoEtAl2025}, each chat message was coded into one of five categories: Category 1, chats that help maintain communication; Category 2, chats that focus discussion to foster progress; Category 3, chats that ask for input on the task; Category 4, chats that contribute details for working through the task; and Category 5, chats that acknowledge partner(s)' input and may continue from that input with their own. We calculated each participant's frequency of messages in the five categories, the total number of messages within each team, and each participant's proportion of the team discussion.

The outcome data contained each participant's final response to each item. Task-level agreement, defined as all four members selecting the same option on all five items, was achieved by 86.0\% of Task Na teams and 93.4\% of Task Nb teams. Item-level agreement ranged from 87.5\% to 89.0\% for Task Na and from 95.5\% to 98.3\% for Task Nb. Thus, agreement was common but not universal, and the common team choice remained latent in the proposed model.

We retained participants who completed at least one negotiation task and teams with exactly four participants assigned to roles P1--P4. The resulting sample included 272 Task Na teams, 289 Task Nb teams, and 1,203 participants. Of these participants, 1,041 completed both tasks (86.5\%), 47 completed only Task Na (3.9\%), and 115 completed only Task Nb (9.6\%).

Chat records were available for 200 Task Na teams and 216 Task Nb teams. At the participant level, 927 participants had process data from at least one task: 800 from Task Na, 864 from Task Nb, and 737 from both tasks. Table~\ref{tab:outcome_process_availability} summarizes the joint availability of outcome and process data. Because team membership differed across tasks, the 737 participants with process data from both tasks did not constitute a complete team-level analytic sample.

\begin{table}[H] 
\centering 
\caption{Participant-level availability of choice outcomes and process data} \label{tab:outcome_process_availability}
\begin{threeparttable} 
\begin{tabular}{llrr} 
\toprule Outcome availability & Process-data availability & $n$ & Percentage \\ 
\midrule 
Na and Nb outcomes & Na and Nb process data & 737 & 61.3\% \\ 
Na and Nb outcomes & No process data & 264 & 21.9\% \\ 
Nb outcome only & Nb process data & 107 & 8.9\% \\ 
Na outcome only & Na process data & 43 & 3.6\% \\ 
Na and Nb outcomes & Na process data only & 20 & 1.7\% \\ 
Na and Nb outcomes & Nb process data only & 20 & 1.7\% \\ 
Nb outcome only & No process data & 8 & 0.7\% \\ 
Na outcome only & No process data & 4 & 0.3\% \\ 
\midrule 
Total & & 1,203 & 100.0\% \\ 
\bottomrule 
\end{tabular} 

\begin{flushleft}
\footnotesize
\textit{Note.} Percentages are calculated relative to the full retained choice-outcome sample of 1,203 participants. The table summarizes data availability at the participant level rather than at the team level. Therefore, participant-level subsets do not necessarily preserve the four-person team structure required by the team-level choice model and team-level process indicators. In addition, Task Na and Task Nb teams were not paired teams with identical members; participants could interact with different teammates across the two tasks. Consequently, retaining participants based only on person-level process availability may break some four-person teams, and retaining complete teams may require including teammates whose process data are not complete for both tasks. 
\end{flushleft} 
\end{threeparttable} 
\end{table}

\paragraph{Imputation of missing process data}

For the latent regression model, process indicators were treated as missing when retained teams lacked matched chat records. Participants who did not send any messages in teams with available chat records were assigned a value of zero rather than treated as missing. Missing values were replaced by the mean of the observed values for the corresponding indicator in the sample used for model estimation. Each indicator was then standardized using the mean and standard deviation from the same sample, so that the imputed values had a standardized value of zero. The baseline and latent regression models used the same response data, whereas the latent regression model additionally incorporated the imputed and standardized process indicators to model the conditional mean of the latent negotiation trait.

\subsection{Results}

\paragraph{Model convergence}

\begin{figure}[H]
    \centering
    \includegraphics[width=0.8\textwidth]{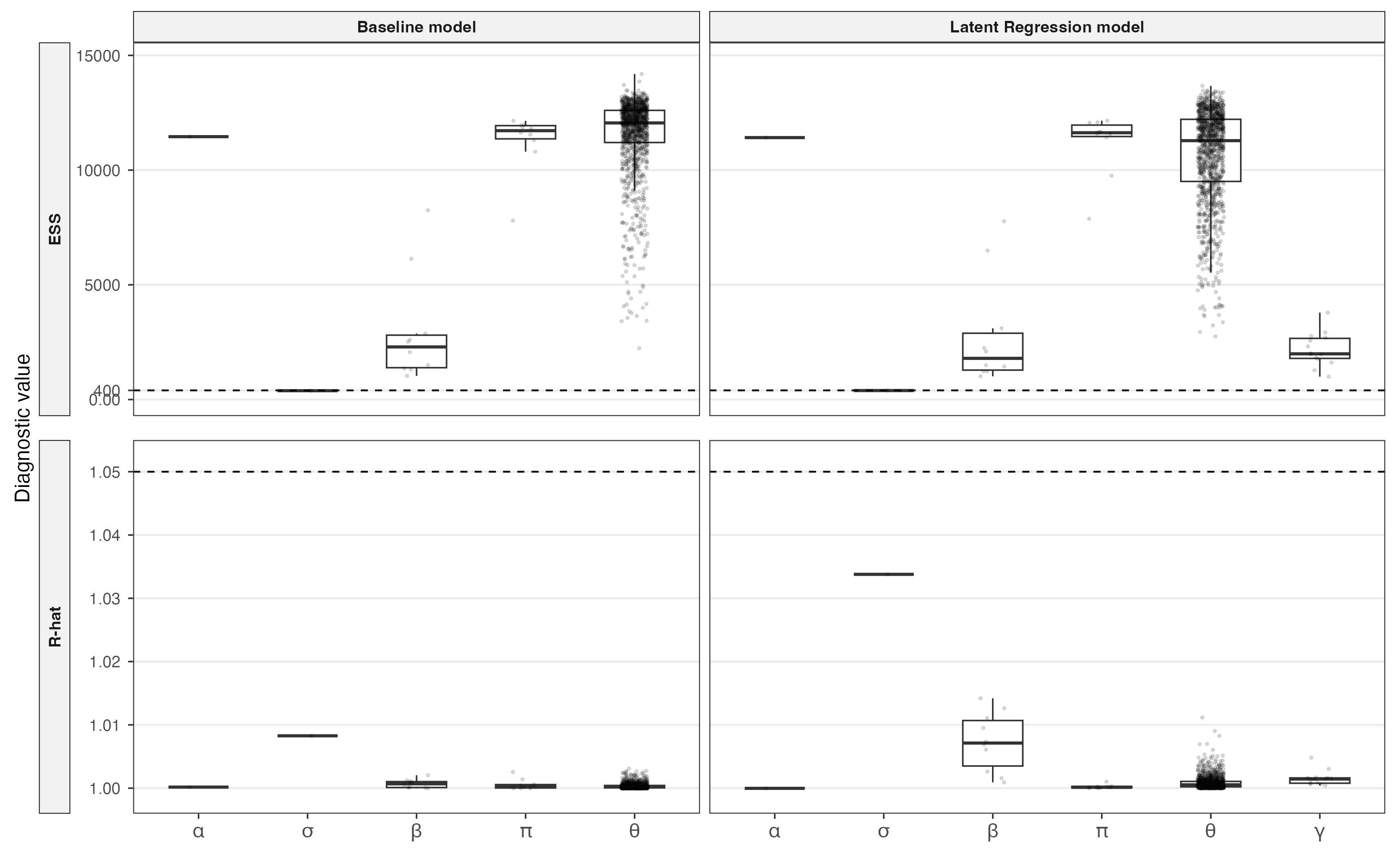}
    \caption{Convergence diagnostics for the baseline and latent regression models.}
    \label{fig:convergence_diagnostics}
\end{figure}

We fitted the baseline and latent regression models to the full dataset. Posterior sampling was based on 40,000 MCMC iterations, with the first 10,000 discarded as burn-in and every second post-burn-in draw retained. Convergence was assessed using the potential scale reduction factor, $\hat{R}$, and the effective sample size (ESS), with $\hat{R} < 1.05$ and ESS $> 400$ used as reference criteria. Figure~\ref{fig:convergence_diagnostics} summarizes the full-data convergence diagnostics for the self-oriented reward-sensitivity parameter ($\alpha$), the item-specific team-choice sensitivity parameters ($\beta_j$), the item-specific self-oriented response probabilities ($\pi_j$), the participant-level latent negotiation traits ($\theta_i$), the latent-trait standard deviation parameter ($\sigma_\theta$), and the latent regression coefficients ($\boldsymbol{\gamma}$). All monitored parameters had $\hat{R} < 1.05$. ESS exceeded 400 for all parameters except $\sigma_\theta$, for which the ESS values were approximately 385 and 392 in the baseline and latent regression models, respectively.

\paragraph{Latent regression effects}

We next examined the estimated effects of the process indicators in the full-data latent regression model. As shown in Table~\ref{tab:process_effects_summary}, self chat proportion had a positive regression coefficient in both tasks. The posterior mean was 0.131 for Task Na, with a 95\% credible interval of $[0.035, 0.229]$, and 0.152 for Task Nb, with a 95\% credible interval of $[0.049, 0.258]$. Thus, participants who contributed a larger proportion of the team discussion tended to have higher latent negotiation traits. The credible intervals for the category-specific communication counts and team total chat all included zero, suggesting that, conditional on participants' relative contribution to the discussion, these additional communication-frequency indicators did not show a clear contribution to variation in the latent negotiation trait.

\begin{table}[htbp]
\centering
\caption{Latent regression effects of process indicators}
\label{tab:process_effects_summary}
\begin{threeparttable}

\begin{tabular}{lrc}
\toprule
Process indicator & Posterior mean & 95\% credible interval \\
\midrule
Na self chat proportion
& 0.131
& $[0.035,\ 0.229]$ \\

Na self count: category 1
& -0.091
& $[-0.220,\ 0.025]$ \\

Na self count: category 2
& -0.087
& $[-0.196,\ 0.011]$ \\

Na self count: category 3
& -0.016
& $[-0.128,\ 0.092]$ \\

Na self count: category 4
& 0.091
& $[-0.046,\ 0.227]$ \\

Na self count: category 5
& 0.029
& $[-0.093,\ 0.151]$ \\

Na team total chat
& -0.036
& $[-0.246,\ 0.175]$ \\

\midrule
Nb self chat proportion
& 0.152
& $[0.049,\ 0.258]$ \\

Nb self count: category 1
& -0.089
& $[-0.205,\ 0.019]$ \\

Nb self count: category 2
& 0.058
& $[-0.029,\ 0.144]$ \\

Nb self count: category 3
& 0.028
& $[-0.078,\ 0.130]$ \\

Nb self count: category 4
& 0.025
& $[-0.114,\ 0.166]$ \\

Nb self count: category 5
& -0.004
& $[-0.123,\ 0.115]$ \\

Nb team total chat
& -0.037
& $[-0.266,\ 0.172]$ \\
\bottomrule
\end{tabular}

\begin{flushleft}
\footnotesize
\textit{Note.}  Posterior means and 95\% credible intervals were obtained from the latent regression model fitted to the full dataset. The coefficients describe associations between the standardized process indicators and the conditional mean of $\log(\theta_i)$. Missing task-specific process indicators were handled using mean imputation before standardization.
\end{flushleft}

\end{threeparttable}
\end{table}

\paragraph{External validity of the estimated latent negotiation trait}

We examined the external validity of the estimated latent negotiation trait using the trait estimates obtained from the models fitted to the full dataset. The analysis included participants who completed both tasks and had available criterion scores. The external criteria included the cognitive score, the STEM score, and the five BFI domain scores \citep{KyllonenEtAl2023,HaoEtAl2025}. Pearson correlations were calculated between these criteria and the latent negotiation traits estimated by the baseline and latent regression models.

As shown in Table~\ref{tab:external_validity_summary}, the estimated latent negotiation traits from both models were positively correlated with the cognitive and STEM scores. For the cognitive score, the correlation was 0.132 for the baseline trait and 0.149 for the latent regression trait. For the STEM score, the corresponding correlations were 0.086 and 0.091. These positive correlations are consistent with the cognitive demands of negotiation, which require participants to evaluate alternative outcomes and their associated rewards when contributing to collective decisions. The slightly higher correlations for the latent regression trait may reflect the additional participant-level information provided by the process indicators when characterizing individual differences in negotiation.

Neither estimated latent negotiation trait was significantly correlated with any of the BFI domains. This pattern suggests that the estimated negotiation trait is not simply a proxy for broad personality characteristics, but captures individual differences that are more specific to participants' influence on collective choice formation.

\begin{table}[H]
\centering
\caption{External validity correlations with cognitive, STEM, and BFI domain measures}
\label{tab:external_validity_summary}
\small
\begin{threeparttable}

\begin{tabular}{lcr@{}lr@{}l}
\toprule
Criterion
& $N$
& \multicolumn{2}{c}{Baseline trait}
& \multicolumn{2}{c}{Latent regression trait}\\
\midrule

Cognitive score
& 1021
& $0.132$ & \textsuperscript{***}
& $0.149$ & \textsuperscript{***}\\

STEM score
& 957
& $0.086$ & \textsuperscript{**}
& $0.091$ & \textsuperscript{**}\\

\midrule

Extraversion
& 957
& $0.024$ &
& $0.045$ &\\

Agreeableness
& 957
& $-0.043$ &
& $-0.036$ &\\

Conscientiousness
& 957
& $0.046$ &
& $0.055$ &\\

Negative emotionality
& 957
& $-0.009$ &
& $-0.021$ &\\

Open-mindedness
& 957
& $0.018$ &
& $0.033$ &\\

\bottomrule
\end{tabular}

\begin{flushleft}
\footnotesize
\textit{Note.} Values are Pearson correlations calculated among participants who completed both tasks and had nonmissing scores on the corresponding external criterion. The baseline and latent regression traits are posterior mean estimates of $\theta_i$ obtained from models fitted to the full dataset. Statistical significance was assessed using conventional two-sided Pearson correlation tests: \textsuperscript{*}$p<.05$, \textsuperscript{**}$p<.01$, and \textsuperscript{***}$p<.001$. These tests assess whether each correlation differs from zero and do not account for dependence induced by team membership.
\end{flushleft}

\end{threeparttable}
\end{table}

\paragraph{Held-out prediction}

We conducted five-fold cross-validation to evaluate the models' out-of-sample performance on the parallel Task Nb, using Task Nb teams as the splitting units. The 289 Task Nb teams were divided into five folds containing 58, 58, 58, 58, and 57 teams. In each fold, all Task Na responses and the Task Nb responses from non-held-out teams were used for model estimation, whereas the responses and process data from the held-out Task Nb teams were excluded. Thus, for participants in the held-out teams who had also completed Task Na, their Task Na responses contributed to model estimation, while their Task Nb responses were reserved for prediction.

To prevent information leakage, the Task Nb process indicators of participants in the held-out teams were treated as missing. Within each fold, missing process indicators were imputed and standardized using only the corresponding estimation sample, following the procedure described above.

Predictive performance was evaluated at both the response and score levels. At the response level, the option with the highest predicted probability was taken as the predicted response. Accuracy was calculated as the proportion of held-out responses for which the predicted and observed options matched. Because each item had five response options, random guessing would yield an expected accuracy of 20\%.

At the score level, the predicted option probabilities were converted into an expected Task Nb total score for each held-out participant who completed both tasks. For each item, the expected score was obtained by weighting each option's reward by its predicted probability and summing across the five options; the resulting expected item scores were then summed across the five Task Nb items. We calculated the Pearson correlation between the expected and observed Task Nb total scores for each model and compared these correlations descriptively with the correlation between the observed Task Na and Task Nb total scores.

As shown in Table~\ref{tab:choice_prediction_accuracy}, both models substantially exceeded the random-guessing benchmark in all five folds, with prediction accuracies ranging from 39.1\% to 43.1\%. The latent regression model had higher accuracy than the baseline model in four of the five folds, with mean accuracies of 42.0\% and 40.8\%, respectively.

\begin{table}[htbp]
\centering
\caption{Held-out Task Nb choice prediction accuracy across five folds}
\label{tab:choice_prediction_accuracy}
\small
\begin{threeparttable}
\begin{tabular}{lrrrrr}
\toprule
Fold & $N_{\text{responses}}$ & $N_{\text{persons}}$ & Baseline & Latent regression & $\Delta$\\
\midrule
1 & 1040 & 208 & .399 & .422 & .023\\
2 & 1025 & 205 & .418 & .421 & .003\\
3 & 1035 & 207 & .417 & .416 & -.001\\
4 & 1045 & 209 & .391 & .410 & .019\\
5 & 1060 & 212 & .415 & .431 & .016\\
\midrule
Mean & -- & -- & .408 (.012) & .420 (.008) & .012 (.010) \\
\bottomrule
\end{tabular}
\begin{flushleft}
\footnotesize
\textit{Note.} $\Delta$ denotes latent regression accuracy minus baseline accuracy. The final row reports the mean and standard deviation across the five folds.
\end{flushleft}
\end{threeparttable}
\end{table}

As shown in Table~\ref{tab:expected_score_correlation}, the mean correlations between the model-based expected and observed Task Nb total scores were .323 for the baseline model and .334 for the latent regression model. Both were slightly higher than the mean correlation of .300 between the observed Task Na and Task Nb total scores, with the latent regression model showing the highest correlation. Overall, the model-based scores showed a modest advantage over the observed Task Na score in predicting Task Nb performance.

\begin{table}[htbp]
\centering
\caption{Correlations between predicted expected Task Nb score and observed Task Nb score}
\label{tab:expected_score_correlation}
\small
\begin{threeparttable}
\begin{tabular}{lrrrr}
\toprule
Fold & $N$ & Baseline & Latent regression & Na score \\
\midrule
1 & 208 & $.464$ & $.464$ & $.420$ \\
2 & 205 & $.312$ & $.349$ & $.250$ \\
3 & 207 & $.433$ & $.451$ & $.361$ \\
4 & 209 & $.186$  & $.167$   & $.234$ \\
5 & 212 & $.197$  & $.214$  & $.222$  \\
\midrule
Mean & -- & .323 (.129) & .334 (.134) & .300 (.088) \\
\bottomrule
\end{tabular}
\begin{flushleft}
\footnotesize
\textit{Note.} The baseline and latent regression columns report correlations between the model-based expected Task Nb scores and the observed Task Nb scores. The Na score column reports correlations between the observed Task Na and Task Nb scores. Mean correlations were obtained by averaging Fisher-transformed correlations across folds and transforming the resulting mean back to the correlation scale. Values in parentheses are standard deviations across the five folds.
\end{flushleft}
\end{threeparttable}
\end{table}

\FloatBarrier

\section{Discussions}

In this paper, we propose a latent choice model for measuring individual negotiation traits from collaborative negotiation data. In the first stage, the latent team choice is modeled using a Luce choice model, in which the probability of selecting an option depends on its team utility and an item-specific team-choice sensitivity parameter. Motivated by item response tree models, the second stage allows each submitted response either to follow the latent team choice or to be generated from the participant's own reward profile. An item-specific self-oriented response probability governs the likelihood of departing from the latent team choice, whereas a global reward-sensitivity parameter determines how strongly options with higher individual rewards are favored when such a departure occurs. The distribution of the negotiation trait can be specified either without participant-level covariates or through a latent regression. The simulation study showed satisfactory parameter recovery across the range of sample sizes and numbers of items considered. Trait recovery improved with both sample size and the number of items, while the estimated traits remained substantially correlated with the true traits under the shorter test conditions. The empirical application used data from two collaborative negotiation tasks, in which participants negotiated in four-person teams and interacted with different teammates across tasks. The main findings include: (1)the estimated negotiation traits showed theoretically consistent associations with external criteria; (2) both models predicted held-out choices well above the accuracy under random guessing and produced expected scores that were positively associated with observed performance on the held-out task; (3) the latent regression model showed modest predictive gains over the baseline model.

We point out that the proposed model has distinct value as a measurement framework for collaborative negotiation. To our knowledge, the proposed model is the first measurement model that uses collective choice data to estimate individual traits in collaborative negotiation tasks. Choice data have a categorical structure similar to the item responses commonly analyzed in item response theory, but responses in collaborative negotiation are jointly generated and therefore dependent across team members. The introduction of a latent team choice accounts for this dependence by linking the members' submitted responses to a common unobserved choice. This formulation makes it possible to estimate individual negotiation traits from final choice data without treating team members' responses as conditionally independent. Beyond accounting for this dependence, the estimated negotiation trait has a direct substantive interpretation. It quantifies the extent to which a participant's reward profile contributes to the team utility underlying latent team choices. Higher trait values strengthen this contribution, making options that yield higher rewards to the participant more likely to be selected. Observed item scores and response agreement, in contrast, summarize outcomes jointly determined by all team members and do not directly indicate individual influence. By combining known reward structures with observations across changing team compositions, the model also provides a basis for studying how different combinations of participants may shape team choices and agreement. The weak associations between the estimated traits and the personality measures further suggest that negotiation influence is not reducible to broad personality characteristics. 

Despite its advantages as a measurement model, the proposed approach has some limitations. First, the interpretation of the negotiation trait relies on repeated observations of participants under different team compositions. In the empirical application, participants completed two parallel tasks with reassigned teammates, allowing recurring individual influence to be distinguished from outcomes specific to a particular team. When the same participants remain together across all items, individual influence may be confounded with persistent team-level patterns, which cannot be separated by the current model without additional structure. This requirement may limit the direct application of the model to settings with fixed teams. Second, missing process indicators were handled using fold-specific mean imputation. This treatment does not account for the mechanism generating the missing data. In the present dataset, unavailable process records may reflect low engagement or invalid task participation rather than random data loss and may therefore be associated with participant characteristics or negotiation behavior. Third, we did not conduct formal significance tests for the differences in predictive performance and external-validity correlations reported in the empirical study. The observed responses were not statistically independent because participants negotiated and responded within teams. Conventional tests that ignore this dependence may underestimate sampling variability and produce overly small \(p\)-values. We therefore interpreted these empirical comparisons descriptively. Future research may use bootstrap or other resampling procedures that preserve the participant- and team-level dependence structure to assess whether the observed differences are statistically significant.

We further discuss several future directions of the proposed method. First, the measurement model may be extended to represent a richer latent negotiation process when the observed data provide sufficient information. The current formulation assumes a single latent team choice for each item. In some negotiations, however, participants with aligned interests may form subgroups, and the influence of a subgroup may not be adequately represented by the sum of its members' individual effects. The model also treats the latent choices for different items separately, although participants may negotiate a set of items jointly by making concessions on some items in exchange for more favorable outcomes on others. In addition, when intermediate or revised choices are recorded before the final submission, the latent negotiation process may be allowed to evolve over time. Sequential latent-state models, such as hidden Markov models, could then be incorporated to represent different stages of negotiation. These extensions would generalize the current two-stage formulation to accommodate more complex negotiation mechanisms.

Second, the structural model may be extended beyond a linear latent regression based on aggregated communication-frequency indicators. Participant characteristics such as gender or socioeconomic status could also be incorporated as covariates alongside process indicators. In addition, richer process information may be relevant to the negotiation trait, including the timing, order, and content of communication. Methods for extracting features from time-stamped process data could be used to preserve the timing and order of the interactions \citep{TangLiuYing2025}, while task-informed sequence-mining methods could be used to construct substantively interpretable behavioral indicators and identify patterns that recur across tasks \citep{HeEtAl2021}. Textual analysis could further provide information about the content of communication. These richer process representations could be incorporated through a more flexible structural model. More fundamentally, rather than first reducing the communication process to participant-level covariates, future work could directly model transitions among communication states or actions conditional on the latent negotiation trait, following sequential response modeling approaches \citep{HanEtAl2022}. Jointly modeling the communication process and negotiation choices may provide a more informative characterization of individual differences in negotiation.

Finally, the model may be extended to measure multiple participant-level latent traits. In the current model, participants differ in their influence on the latent team choice, whereas self-oriented deviations at the response stage are governed by item-specific and global parameters that are shared across participants. A multidimensional extension could introduce a separate latent trait for the tendency to follow or depart from the latent team choice. This extension would distinguish individual influence on collective choice formation from self-oriented behavior in the submitted responses. Additional dimensions could also be considered when the assessment design includes items intended to elicit different negotiation mechanisms and provides sufficient information to identify these dimensions.

\section*{Acknowledgements}

During the preparation of this manuscript, the authors used ChatGPT (OpenAI) to assist with improving the clarity, grammar, and readability of the text and to improve code readability and documentation. The authors reviewed, edited, and verified all content generated with the assistance of this tool and take full responsibility for the accuracy and integrity of the work.

\section*{Data availability statement}

The empirical data analyzed in this study are subject to data access restrictions and cannot be redistributed by the authors. The code is available in an anonymized repository at \href{https://osf.io/pdge5/overview?view_only=dd77978333da4fbfb33fc8c752dd147d}{anonymized OSF repository}.

\begin{Backmatter}

\vspace{1cm}
\printbibliography

@article{MartinRaughEtAl2020,
  author  = {Martin-Raugh, Michelle P. and Kyllonen, Patrick C. and
             Hao, Jiangang and Bacall, Adam and Becker, Dovid and
             Kurzum, Christopher and Yang, Zhitong and Yan, Fred and
             Barnwell, Patrick},
  title   = {Negotiation as an Interpersonal Skill: Generalizability of
             Negotiation Outcomes and Tactics across Contexts at the
             Individual and Collective Levels},
  journal = {Computers in Human Behavior},
  year    = {2020},
  volume  = {104},
  pages   = {105966},
  doi     = {10.1016/j.chb.2019.03.030}
}

@article{JiangEtAl2023,
  author  = {Jiang, Yang and Mart{\'i}n-Raugh, Michelle and Yang, Zhitong and
             Hao, Jiangang and Liu, Lei and Kyllonen, Patrick C.},
  title   = {Do You Know Your Partner's Personality through Virtual
             Collaboration or Negotiation? Investigating Perceptions of
             Personality and Their Impacts on Performance},
  journal = {Computers in Human Behavior},
  year    = {2023},
  volume  = {141},
  pages   = {107608},
  doi     = {10.1016/j.chb.2022.107608}
}

@article{ThompsonEtAl2010,
  author  = {Thompson, Leigh L. and Wang, Jiunwen and Gunia, Brian C.},
  title   = {Negotiation},
  journal = {Annual Review of Psychology},
  year    = {2010},
  volume  = {61},
  pages   = {491--515},
  doi     = {10.1146/annurev.psych.093008.100458}
}

@article{BoothbyEtAl2023,
  author  = {Boothby, Erica J. and Cooney, Gus and
             Schweitzer, Maurice E.},
  title   = {Embracing Complexity: {A} Review of Negotiation Research},
  journal = {Annual Review of Psychology},
  year    = {2023},
  volume  = {74},
  pages   = {299--332},
  doi     = {10.1146/annurev-psych-033020-014116}
}

@article{ElfenbeinEtAl2008,
  author  = {Elfenbein, Hillary Anger and Curhan, Jared R. and
             Eisenkraft, Noah and Shirako, Aiwa and Baccaro, Lucio},
  title   = {Are Some Negotiators Better Than Others?
             {Individual} Differences in Bargaining Outcomes},
  journal = {Journal of Research in Personality},
  year    = {2008},
  volume  = {42},
  number  = {6},
  pages   = {1463--1475},
  doi     = {10.1016/j.jrp.2008.06.010}
}

@article{Thurstone1927,
  author  = {Thurstone, L. L.},
  title   = {A Law of Comparative Judgment},
  journal = {Psychological Review},
  year    = {1927},
  volume  = {34},
  number  = {4},
  pages   = {273--286},
  doi     = {10.1037/h0070288}
}

@book{Luce1959,
  author    = {Luce, R. Duncan},
  title     = {Individual Choice Behavior: A Theoretical Analysis},
  publisher = {John Wiley \& Sons},
  address   = {New York},
  year      = {1959}
}

@article{Plackett1975,
  author  = {Plackett, R. L.},
  title   = {The Analysis of Permutations},
  journal = {Journal of the Royal Statistical Society:
             Series C (Applied Statistics)},
  year    = {1975},
  volume  = {24},
  number  = {2},
  pages   = {193--202},
  doi     = {10.2307/2346567}
}

@article{MaydeuOlivaresBockenholt2005,
  author  = {Maydeu-Olivares, Albert and B{\"o}ckenholt, Ulf},
  title   = {Structural Equation Modeling of Paired-Comparison
             and Ranking Data},
  journal = {Psychological Methods},
  year    = {2005},
  volume  = {10},
  number  = {3},
  pages   = {285--304},
  doi     = {10.1037/1082-989X.10.3.285}
}

@article{MaydeuOlivaresBrown2010,
  author  = {Maydeu-Olivares, Alberto and Brown, Anna},
  title   = {Item Response Modeling of Paired Comparison
             and Ranking Data},
  journal = {Multivariate Behavioral Research},
  year    = {2010},
  volume  = {45},
  number  = {6},
  pages   = {935--974},
  doi     = {10.1080/00273171.2010.531231}
}

@article{BrownMaydeuOlivares2011,
  author  = {Brown, Anna and Maydeu-Olivares, Alberto},
  title   = {Item Response Modeling of Forced-Choice Questionnaires},
  journal = {Educational and Psychological Measurement},
  year    = {2011},
  volume  = {71},
  number  = {3},
  pages   = {460--502},
  doi     = {10.1177/0013164410375112}
}

@article{DeBoeckPartchev2012,
  author  = {De Boeck, Paul and Partchev, Ivailo},
  title   = {{IRTrees}: Tree-Based Item Response Models
             of the {GLMM} Family},
  journal = {Journal of Statistical Software},
  year    = {2012},
  volume  = {48},
  number  = {1},
  pages   = {1--28},
  doi     = {10.18637/jss.v048.c01}
}

@article{JeonDeBoeck2016,
  author  = {Jeon, Minjeong and De Boeck, Paul},
  title   = {A Generalized Item Response Tree Model
             for Psychological Assessments},
  journal = {Behavior Research Methods},
  year    = {2016},
  volume  = {48},
  number  = {3},
  pages   = {1070--1085},
  doi     = {10.3758/s13428-015-0631-y}
}

@article{ZhangKuwanoEtAl2009,
  author  = {Zhang, Junyi and Kuwano, Masashi and Lee, Backjin
             and Fujiwara, Akimasa},
  title   = {Modeling Household Discrete Choice Behavior Incorporating
             Heterogeneous Group Decision-Making Mechanisms},
  journal = {Transportation Research Part B: Methodological},
  year    = {2009},
  volume  = {43},
  number  = {2},
  pages   = {230--250},
  doi     = {10.1016/j.trb.2008.05.002}
}

@article{ChiapporiMazzocco2017,
  author  = {Chiappori, Pierre-Andre and Mazzocco, Maurizio},
  title   = {Static and Intertemporal Household Decisions},
  journal = {Journal of Economic Literature},
  year    = {2017},
  volume  = {55},
  number  = {3},
  pages   = {985--1045},
  doi     = {10.1257/jel.20150715}
}

@techreport{KyllonenEtAl2023,
  author      = {Kyllonen, Patrick and Hao, Jiangang and Weeks, Jonathan and Fauss, Michael and Kerzabi, Emily},
  title       = {Collaborative Problem Solving ({CPS}) Skill: Estimating an Individual's Contribution to Small Group Performance},
  year        = {2023},
  note   = {Unpublished manuscript, Educational Testing Service}
}

@article{HaoEtAl2025,
  author  = {Hao, Jiangang and Cui, Wenju and Kyllonen, Patrick C. and
             Kerzabi, Emily and Liu, Lei and Flor, Michael},
  title   = {Automated coding of communications in collaborative
             problem-solving tasks using {ChatGPT}},
  journal = {Journal of Educational Measurement},
  year    = {2025},
  volume  = {62},
  number  = {4},
  pages   = {809--837},
  doi     = {10.1111/jedm.70014}
}

@article{TangLiuYing2025,
  author  = {Tang, Xueying and Liu, Jingchen and Ying, Zhiliang},
  title   = {A path signature perspective of process data feature extraction},
  journal = {British Journal of Mathematical and Statistical Psychology},
  year    = {2025},
  volume  = {78},
  number  = {3},
  pages   = {939--964},
  doi     = {10.1111/bmsp.12390}
}

@article{HeEtAl2021,
    author = {He, Qiwei and Borgonovi, Francesca and Paccagnella, Marco},
	title = {Leveraging process data to assess adults’ problem-solving skills: {Using} sequence mining to identify behavioral patterns across digital tasks},
    journal = {Computers \& Education},
    year = {2021},
	volume = {166},
	pages = {104170},
	doi = {10.1016/j.compedu.2021.104170},
}

@article{EstherEtAl2022,
	author = {Ulitzsch, Esther and Pohl, Steffi and Khorramdel, Lale and Kroehne, Ulf and von Davier, Matthias},
    title = {A {Response}-{Time}-{Based} {Latent} {Response} {Mixture} {Model} for {Identifying} and {Modeling} {Careless} and {Insufficient} {Effort} {Responding} in {Survey} {Data}},
	journal = {Psychometrika},
    year = {2022},
    volume = {87},
    pages = {593--619},
	doi = {10.1007/s11336-021-09817-7},
}

@article{EstherEtAl2020,
    author = {Ulitzsch, Esther and von Davier, Matthias and Pohl, Steffi},
	title = {A hierarchical latent response model for inferences about examinee engagement in terms of guessing and item‐level non‐response},
    journal = {British Journal of Mathematical and Statistical Psychology},
    year = {2020},
	volume = {73},
	pages = {83--112},
	doi = {10.1111/bmsp.12188},
}

@article{HanEtAl2022,
    author = {Han, Yuting and Liu, Hongyun and Ji, Feng},
	title = {A {Sequential} {Response} {Model} for {Analyzing} {Process} {Data} on {Technology}-{Based} {Problem}-{Solving} {Tasks}},
	volume = {57},
	journal = {Multivariate Behavioral Research},
    year = {2022},
	pages = {960--977},
	doi = {10.1080/00273171.2021.1932403},
}

@article{AnderssonXin2021,
    author = {Andersson, Björn and Xin, Tao},
	title = {Estimation of {Latent} {Regression} {Item} {Response} {Theory} {Models} {Using} a {Second}-{Order} {Laplace} {Approximation}},
    volume = {46},
    journal = {Journal of Educational and Behavioral Statistics},
    year = {2021},
	pages = {244--265},
	doi = {10.3102/1076998620945199},
}

@article{Mislevy1985,
    author = {Robert J. Mislevy},
    title = {Estimation of Latent Group Effects},
    volume = {80},
    journal = {Journal of the American Statistical Association},
    year = {1985},
    pages = {993--997},
    doi = {10.2307/2288565}
}

\end{Backmatter}

\clearpage

\appendix

\section*{Appendix}

\setcounter{table}{0}
\renewcommand{\thetable}{A\arabic{table}}

\section{Estimation Details}
\label{app:estimation}

The prior distribution of the participant-level latent negotiation trait $\theta_i$ is specified through the corresponding structural model described in the main text. The prior distributions for the remaining model parameters are summarized in Table~\ref{tab:prior-specifications}. The same prior specifications were used in the simulation and empirical studies.

\begin{table}[htbp]
\centering
\caption{Prior specifications}
\label{tab:prior-specifications}
\small
\begin{threeparttable}
\begin{tabular}{ll}
\toprule
Parameter & Prior distribution \\
\midrule
$\beta_j$ 
    & $\log(\beta_j) \sim N(0.50,\,0.50^2)$ \\
$\pi_j$ 
    & $\pi_j \sim \operatorname{Uniform}(0,1)$ \\
$\alpha$ 
    & $\log(\alpha) \sim N(\log 5,\,0.30)$ \\
$\boldsymbol{\gamma}$
    & $\boldsymbol{\gamma} \sim N(\mathbf{0},\mathbf{I})$ \\
$\sigma_\theta$ 
    & $\sigma_\theta \sim \operatorname{Half\text{-}Normal}(0,1)$ \\
\bottomrule
\end{tabular}
\begin{flushleft}
\footnotesize
\textit{Note.} The prior for $\boldsymbol{\gamma}$ applies only to the latent regression model.
\end{flushleft}
\end{threeparttable}
\end{table}

Posterior sampling was conducted using two Markov chains. For the simulation study, each chain was run for 20,000 iterations, with the first 5,000 discarded as burn-in and the remaining draws thinned by retaining every second draw. For the empirical analysis, each chain was run for 40,000 iterations, with the first 10,000 discarded as burn-in and the remaining draws thinned in the same manner. Random-walk Metropolis--Hastings updates were used for the nonconjugate parameters $\alpha$, $\boldsymbol{\beta}$, $\boldsymbol{\pi}$, and $\boldsymbol{\theta}$, with $\alpha$, $\boldsymbol{\beta}$, and $\boldsymbol{\theta}$ sampled on the log scale to ensure positivity. Conjugate updates were used for $\boldsymbol{\gamma}$ in the latent regression model. For the empirical analysis, a slice sampler was used for $\sigma_\theta$ to
improve sampling efficiency. Convergence was assessed using $\hat{R}$ and the effective sample size (ESS).

\end{document}